\documentclass[a4paper]{article}
\usepackage{qcircuit}
\usepackage{graphicx}
\usepackage{amsmath}

\usepackage[utf8x]{inputenc}
\usepackage[T1]{fontenc}
\usepackage{authblk}
\usepackage{subfig}

\usepackage[a4paper,top=3cm,bottom=2cm,left=3cm,right=3cm,marginparwidth=1.75cm]{geometry}
\input{amssym}

\usepackage[colorinlistoftodos]{todonotes}
\usepackage[colorlinks=true, allcolors=blue]{hyperref}
\usepackage{authblk}

\newcommand{\ket}[1]{| #1 \rangle}
\newcommand{\bra}[1]{\langle #1 |}

\title{\bf Hybrid quantum variational algorithm for simulating open quantum systems with near-term devices }

\author[1]{Mahmoud Mahdian $^{1}$,  H.Davoodi Yeganeh$^1$}
\date{%
    $^1$Faculty of Physics, Theoretical and astrophysics department , University of Tabriz, 51665-163 Tabriz, Iran\\
       \today}

\begin{document}
\maketitle

\begin{abstract}

Hybrid quantum-classical (HQC) algorithms make it possible to use near-term quantum devices supported by classical computational resources by useful control schemes. In this paper, we develop an HQC algorithm using an efficient variational optimization approach to simulate open system dynamics under the Noisy-Intermediate Scale Quantum(NISQ) computer. Using the time-dependent variational principle (TDVP) method and extending it to McLachlan TDVP for density matrix which involves minimization of Frobenius norm of the error, we apply the unitary quantum circuit to obtain the time evolution of the open quantum system in the Lindblad formalism. Finally, we illustrate the use of our methods with detailed examples which are in good agreement with analytical calculations.

\end{abstract}


\section{Introduction}
As we know, real quantum systems are not isolated from their surroundings, but always interact with the environmental degrees of freedom as an open quantum system. These interactions generally lead to dissipation and decoherence, affecting the behavior of the system. Due to the type of interaction and strength of the coupling between the environment and the system, the dynamics of an open quantum system can be divided into two categories; Markovian and non-Markovian. Markovian dynamics (memoryless), which we will review in this paper, can be described by the first-order differential equation when the system interacts weekly with its surroundings and modeled utilizing a dynamical semigroup in Lindblad form. But in non-Markovian dynamics, the memory effects cannot be ignored, and in this case, the information enters the environment and backflow from it. So considering the dynamics of open systems with many degrees of freedom is one of the significant challenges and allows us to a better understanding of the nonequilibrium dynamics of many-body quantum systems. There are well-known numerical and analytical methods for calculating the dynamics of quantum systems, e.g. Time-dependent variational quantum Monte Carlo \cite{Carleo}, Nonequilibrium dynamical mean-field theory \cite{Aoki}, Time-dependent density-matrix renormalization-group \cite{Daley} , Hierarchy equations of motion (HEOM)\cite{Ishizaki} and Multi-configuration time-dependent Hartree\cite{Wang}. However, with the increase in the size of the system, these classical methods are inefficient and take exponential time .\\
Hybrid variational quantum-classical algorithms are consisting of a relatively low-depth quantum circuit, designed to utilize both quantum and classical resources to solve specific tasks not accessible to traditional classical computers. This method applies to the quantum simulation of many-body systems for a variety of complex problems including, dynamics\cite{Li2016EfficientMinimisation}, optimization \cite{Kandala, Peruzzo, Farhi, Farhi2}, quantum chemistry\cite{Alexander, Yudong} and so on. The significant benefit of this method is that it gives rise to a setup that can have much less strict hardware requirements and promising for the near-term application of small quantum computing. The main idea of this method is dividing the problem into two parts that each of performing a single task and can be implemented efficiently on a classical and a quantum computer. Theoretically and experimentally, it has been shown that using near-term quantum devices containing from tens to hundreds of qubits and limited error correction can achieve quantum supremacy. But it is not enough for quantum algorithms to perform more complex calculations in large-scale quantum systems, where HQC algorithms are the best candidate with low sequences of quantum gate operations. Recently, several HQC methods for solving specific optimization tasks have been developed. Two representative variational HQC are the variations quantum eigensolver (VQE) which is a hybrid algorithm to approximate the ground state eigenvalues for chemical systems \cite{Kandala, Peruzzo} and the quantum approximate optimization algorithm (QAOA) for finding an approximate solution of an optimization problem\cite{Farhi, Farhi2}.\\
One of the conventional methods used in most quantum algorithms to simulate the dynamics of quantum systems is commonly referred to as Trotterization, where the Hamiltonian can be decomposed into a sequence of quantum gates and requires many operations to step through the time evolution, e.g., in the single time step, $O(N)$ gates and $O(\sqrt{N})$ depth\cite{Masuo1, Masuo2}. Another way to calculate system dynamics is based on variational technique, which first assumes the trial state that depends on the controllable parameters, and then we evaluate a state that provides a excellent approximation to the real state by changing the parameters \cite{Kreula}. \\
As mentioned, the dynamics of an open system can be described by the master equation, which contains a unitary part under the system's Hamiltonian and non-unitary components leading to the decoherence and dissipative. In recent years, various efforts have been developed to simulate the evolution of open quantum system developments \cite{Wei11, Zixuan, Yoshioka}. Most of these papers require auxiliary qubits for simulation of open systems due to the non-unitary part of dynamics, which is the effect of the system's interaction with the environment. One of the effective methods to perform time evaluation of many-body systems is the time-dependent variational principle (TDVP), which is obtained from the Dirac-Frenkel variational principle \cite{Dirac11, McLachlan, Langhoff}. This method has many applications in quantum chemistry based on the time-dependent Hartree-Fock theory \cite{Arrighini} and strongly correlated systems. Recently, Doriol et al. extended the McLachlan TDVP method to the density matrix for open quantum systems based on the Lindblad approach \cite{Doriol}. In this paper, we described and analyzed an efficient manner to realize a plan for simulating open system dynamics that were base on the HQC algorithm. This system depends on the TDVP to optimal gradient control iterative that combines the classical and quantum processors . We design a quantum algorithm for the unitary and non-unitary parts of the Markovian dynamics via Liouvillian superoperator by minimization of the Frobenius norm as a cost function up to a small error. We study our method for obtaining the dynamics of open quantum systems in the presence of different quantum noise, which is in good agreement with analytical calculations. The essential constituent of our simulation algorithm is the potentiality to simulate dynamics of open systems on a quantum computer, and it has found critical applications for a great variety of computational tasks, such as nuclear physics, simulating condensed-matter systems, calculating molecular properties, chemical reaction dynamics\cite{IvanKassal}, and probing quantum effects in biological systems \cite{NeillLambert}.\\
The paper is organized as follows. In Sec. 2, we briefly introduce the TDVP method for density matrix, and then we propose a hybrid quantum algorithm for simulating the dynamics of an open quantum system. In Sec. 3, presents our numerical results. Finally, Sec. 4 gives the conclusions with further discussion on future works.

\section{Our Methods}
The variational method was first introduced by Dirac \cite{Dirac11} and Frenkel \cite{Frenkel} in 1930 to approximate the time-dependent Schrödinger equation. In this method, the wavefunction of the system is determined by a set of parameters and the equation of motion of these parameters is obtained using the variational principle. This method has significant applications in hybrid quantum algorithms; e.g., in Ref \cite{Li2018}, the authors introduced a hybrid quantum algorithm for simulating quantum dynamics of closed systems based on variational principle, which can be implemented on NISQ devices. In our approach, we will extend the HQC algorithm to the dynamics of open quantum systems.

\subsection{Time-dependent variational methods}
The master equation can be described by $\dot{\rho}(t)=\mathcal{L}(\rho)$, which is a linear and time-local equation for the evaluation of the density matrix in $d-dimensional$ Hilbert space. $\mathcal{L}$ is then an linear map usually called Liouvillian superoperator that contains both the unitary and the non-unitary part which results from the interaction between the system and the environment. In this section, we consider evolution of a quantum density matrix $\rho(t)$ of the form
 \begin{equation}\label{eq:rho_t}
 \rho(t)=\sum_{j=1}^N\sum_{k=1}^N B_{jk}(t)\ket{\psi_j(t)}\bra{\psi_k(t)}
 \end{equation}
under the Lindbladian operator
\begin{equation}\label{eq:L_rho}
\mathcal{L}(\rho)=-i[H,\rho]+\sum_{j=1}^{K}\gamma_j(L_j\rho L_j^\dagger-\frac{1}{2}\{L_j^\dagger L_j,\rho\}).
\end{equation}
Here, we maintain the $N^2$ coefficients of $B_{jk}$ in a matrix $\bf B$ and the pure states $\{|\psi_k(t)\rangle\}_{k=1}^N$ are each prepared with a parametrized quantum circuit. For each $k$, we have
\begin{equation}\label{eq:psi_kprep}
|\psi_k(t)\rangle = U_k(t)|0\rangle = \prod_{\alpha}U_{k\alpha}(z_{k\alpha}(t))|0\rangle
\end{equation}
where $z_{k\alpha}(t)$ is the $\alpha$-th parameter in the circuit generating $|\psi_k(t)\rangle$. Note that the states $\{|\psi_k(t)\rangle\}_{k=1}^N$ may not necessarily be orthogonal.
The general settings of the HQC algorithm has four stages:
\begin{enumerate}
\item Each unitary $U_{k\alpha}$ in \eqref{eq:psi_kprep} takes the form $\exp(-iz_{k\alpha}H_{k\alpha})$ and can be implemented in $O(1)$ time. Each Hamiltonian term $H_{k\alpha}$ is a sum of $O(1)$ terms which are tensor products of Pauli operators.
\item Each parametrized circuit for preparing  the states $\{|\psi_k(t)\rangle\}$ contains at most $m$ unitary operators $U_{k\alpha}$.
\item The time dependence of $z_{k\alpha}$ should be such that $|\frac{\partial z_{k\alpha}}{\partial t}|$ is bounded by a constant.
\item The Hamiltonian $H$ in $\mathcal{L}$ can be rewritten as a linear combination of $r$ tensor products of Pauli operators. Each Lindblad jump operator $L_j$ can be expressed as a linear combination of $s$ tensor products of Pauli operators.
\end{enumerate}

\subsubsection{Simulation dynamics via outer-product method}
The basic idea of the simulation is that given the state at current time $t$, determine the state $t+\Delta t$ according to time-dependent variational principle \cite{McLachlan}, which dictates that the Frobenius norm $\|\dot\rho-\mathcal{L}(\rho)\|$ must be minimized. This translates to the stationary condition $\text{Tr}\{\delta\rho(\dot\rho-\mathcal{L}(\rho))\}=0$, which leads to the Equation of motion for updating the matrix $\bf B$ and the parameters $\{z_{k\alpha}\}$ \cite{Joubert-Doriol2015Problem-freeSystems}:

\begin{equation}\label{eq:eom}
\begin{array}{ccl}
\dot{\bf B} & = & {\bf \mathcal{S}}^{-1}{\bf L}{\bf  \mathcal{S}}^{-1}-({\bf  \mathcal{S}}^{-1}\boldsymbol\tau{\bf B}+{\bf B}\boldsymbol\tau^\dagger{\bf \mathcal{S}}^{-1}) \\
\dot{\bf z} & = & {\bf  \mathcal{C}}^{-1}{\bf  \mathcal{Y}}
\end{array}
\end{equation}
where $\bf \mathcal{S}$ is the overlap matrix such that $ \mathcal{S}_{jk}=\langle\psi_j|\psi_k\rangle$. Here we omit the time dependence of the objects for notational simplicity. The operator $\bf L$ is such that $L_{jk}=\langle\psi_j|\mathcal{L}(\rho)|\psi_k\rangle$. $\boldsymbol\tau$ is a matrix with $\tau_{jk}=\langle \psi_j|\dot\psi_k\rangle$. The second equation in \eqref{eq:eom} is a linear system for updating the circuit parameters which generate the pure states $\{|\psi_k\rangle\}$. $\bf z$ is all $\{z_{k\alpha}\}$ stored in a vector and the elements of vector $\bf  \mathcal{Y}$ are also indexed by two indices. Each element of the matrix $\bf  \mathcal{C}$, accordingly, is indexed by four indices. From \cite{Joubert-Doriol2015Problem-freeSystems} we have
\begin{equation}\label{eq:C}
\begin{array}{ccl}
 \mathcal{C}_{k\alpha,\ell\beta} & = & \displaystyle \left[\left\langle\frac{\partial\psi_k}{\partial z_{k\alpha}}\left\vert\frac{\partial\psi_\ell}{\partial z_{\ell\beta}}\right.\right\rangle-\sum_{p,q}[{\bf \mathcal{S}}^{-1}]_{pq}\left\langle\left.\frac{\partial\psi_k}{\partial z_{k\alpha}}\right|\psi_p\right\rangle\left\langle\psi_q\left|\frac{\partial\psi_\ell}{\partial z_{\ell\beta}}\right.\right\rangle\right]\cdot[{\bf B}{\bf \mathcal{S}}{\bf B}]_{\ell k} \\[0.1in]
 \mathcal{Y}_{k\alpha} & = & \displaystyle \sum_\ell\left[\left\langle\frac{\partial\psi_k}{\partial z_{k\alpha}}|\mathcal{L}(\rho)|\psi_\ell\right\rangle-\sum_{p,q}[{\bf  \mathcal{S}}^{-1}]_{pq}\left\langle\left.\frac{\partial\psi_k}{\partial z_{k\alpha}}\right|\psi_p\right\rangle\langle\psi_q|\mathcal{L}(\rho)|\psi_\ell\rangle\right]\cdot B_{\ell k}.
\end{array}
\end{equation}
To estimate the resource needed for implementing the simulation scheme, we consider both the cost of computing the objects in Equation \ref{eq:eom} as well as the cost of solving Equation \ref{eq:eom}. The latter cost can be estimated as $O(N^3m^3)$ since the matrix $\bf  \mathcal{C}$ is $(Nm)\times (Nm)$ and the cost of inverting $\bf  \mathcal{C}$ dominates over that of the equation of motion for $\bf B$. The cubic scaling assumes Gaussian elimination for solving the linear systems. We then proceed by considering the resource needed for computing each of the quantities on the right-hand side of Equation \ref{eq:eom}.

$\quad$\\
\noindent{\bf Compute ${\bf  \mathcal{S}}^{-1}$.} Each element of $\bf  \mathcal{S}$ is an overlap $\langle \psi_j|\psi_k\rangle = \langle 0|U_j^\dagger U_k|0\rangle$. Since the circuits for preparing each pure state $|\psi_k(t)\rangle$ is known, one could evaluate the real and imaginary parts of the overlap $\langle\psi_k(t)|\psi_j(t)\rangle$ with simple quantum circuits
\cite{Buhrman2001QuantumFingerprinting,Chamorro-Posada2017TheEvolutions}.
The cost for evaluating each element of $\bf  \mathcal{S}$ up to error $\epsilon$ is then $O(m/\epsilon^2)$, leading to a cost for computing the entire $\bf \mathcal{S}$ to be $O(N^2m/\epsilon^2)$. Inverting $\bf  \mathcal{S}$ costs $O(N^3)$ via Gaussian elimination. Hence the total cost of computing ${\bf \mathcal{S}}^{-1}$ is $O(N^3+\frac{N^2m}{\epsilon^2})$.

$\quad$\\
\noindent{\bf Compute $\boldsymbol\tau$}. Each element of $\boldsymbol\tau$ is $\tau_{k\ell}=\langle\psi_k|\dot\psi_\ell\rangle$. Here $|\dot\psi_\ell\rangle$ is the time derivate of the state $|\psi_\ell\rangle$. More explicitly,
\begin{equation}
|\dot\psi_\ell\rangle = \sum_\alpha\left[\cdots U_{\ell(\alpha+1)}\frac{\partial U_{\alpha\ell}}{\partial z_{\ell\alpha}}\frac{\partial z_{\ell\alpha}}{\partial t}U_{\ell(\alpha-1)}\cdots\right]|0\rangle.
\end{equation}
Note that $\frac{\partial z_{\ell\alpha}}{\partial t}$ is a scalar and $\frac{\partial U_{\ell\alpha}}{\partial z_{\ell\alpha}}=-iH_{\ell\alpha}U_{\ell\alpha}$ from the general settings discussed above. Hence we have
\begin{equation}\label{eq:tau_kl}
\tau_{k\ell}=\langle\psi_k|\dot\psi_\ell\rangle = -i\sum_\alpha\frac{\partial z_{\ell\alpha}}{\partial t}
\underbrace{
\langle 0|U_k^\dagger\left[\cdots U_{\ell(\alpha+1)}H_{\ell\alpha}U_{\ell\alpha}U_{\ell(\alpha-1)}\cdots\right]|0\rangle}.
\end{equation}
Here each term in the bracket can be estimated by first decomposing $H_{\ell\alpha}$ into tensor products of Pauli operators. From the general setting there are at most $O(1)$ number of those terms. Each of such Pauli terms can be estimated with circuits of the same form as those computing $S_{jk}$, which costs $O(m/\epsilon^2)$ to estimate up to error $\epsilon$. Hence each term in the bracket in \eqref{eq:tau_kl} costs $O(m/\epsilon^2)$. Since estimating a sum of $M$ terms with variance $O(1)$ to error $\epsilon$ requires $O(M^2/\epsilon^2)$ state preparation and measurements \cite{McClean2016TheAlgorithms}, to compute $\tau_{k\ell}$ costs in total $O(m^3/\epsilon^2)$. Computing the entire $\boldsymbol\tau$ therefore costs $O(N^2m^3/\epsilon^2)$.

$\quad$\\
\noindent{\bf Compute $\bf L$.} Each element of $\bf L$ is $L_{k\ell}=\langle\psi_k|\mathcal{L}[\rho]|\psi_\ell\rangle$. Substituting Equations \ref{eq:rho_t} and \ref{eq:L_rho} into the definition of $L_{k\ell}$, we have
\begin{equation}
\begin{array}{ccl}
L_{k\ell} & = & \displaystyle -i\sum_{p,q}B_{pq}\left( \mathcal{S}_{q\ell}\langle\psi_k|H|\psi_p\rangle- \mathcal{S}_{kp}\langle\psi_q|H|\psi_\ell\rangle\right) \\[0.2in]
 & - & \displaystyle \sum_j\gamma_j\sum_{p,q}B_{pq}\left[\langle\psi_k|L_j|\psi_p\rangle\langle\psi_q|L_j^\dagger|\psi_\ell\rangle-\frac{1}{2}\left( \mathcal{S}_{q\ell}\langle\psi_k|L_j^\dagger L_j|\psi_p\rangle
 + \mathcal{S}_{kp}\langle\psi_q|L_j^\dagger L_j|\psi_l\rangle
 \right)\right].
\end{array}
\end{equation}
Note that $L_{k\ell}$ is computed from three types of terms: 1) $\langle\psi_j|H|\psi_k\rangle$, which costs in total $O(N^2mr^2/\epsilon^2)$ measurements to estimate to error $\epsilon$; 2) $\langle\psi_k|L_j|\psi_p\rangle$, which costs in total $O(N^2ms^2/\epsilon^2)$ measurements to estimate within error $\epsilon$; 3) $\langle\psi_k|L_j^\dagger L_j|\psi_\ell\rangle$ terms, which costs in total $O(N^2ms^4/\epsilon^2)$ to estimate within error $\epsilon$. Once these three sets of terms are computed, all matrix elements $L_{k\ell}$ can be computed easily. Since each $L_{k\ell}$ contains $O(N^2s)$ terms, the cost of computing the three types of terms is $O(N^6ms^2(r^2+s^4)/\epsilon^2)$ to guarantee variance in estimating $L_{k\ell}$ under $\epsilon^2$. Once these three types of terms are computed, putting together the matrix ${\bf L}$ costs $O(N^4s)$ time, which is of much lower order. Therefore the cost of computing ${\bf L}$ is $O(N^6ms^2(r^2+s^4)/\epsilon^2)$.

$\quad$\\
\noindent{\bf Computing $\bf  \mathcal{C}$}. From Equation \eqref{eq:C} we see that there are two sets of terms that need to be evaluated:
\begin{enumerate}
\item Overlap terms of the form $\langle\frac{\partial\psi_k}{\partial z_{k\alpha}}|\frac{\partial \psi_\ell}{\partial z_{\ell\beta}}\rangle$. Clearly if $H_{k\alpha}$ and $H_{\ell\beta}$ can be expanded into $O(1)$ terms which are tensor products of Pauli operators, the overlap term consists of $O(1)$ measurable quantities. Each quantity requires $O(m/\epsilon^2)$ measurements to estimate up to error $\epsilon$. Then estimating a single overlap costs $O(m/\epsilon^2)$, leading to the total cost of estimating the overlap terms $O(N^2m^3/\epsilon^2)$.
\item Overlap terms of the form $\langle\frac{\partial \psi_k}{\partial z_{k\alpha}}|\psi_p\rangle$. The analysis is similar to above, except that now there are only $O(N^2m)$ such terms to evaluate, leading to a total cost of $O(N^2m^2/\epsilon^2)$.
\end{enumerate}
Because each $ \mathcal{C}_{k\alpha,\ell\beta}$ element contains $O(N^2)$ terms, the total cost of estimating the matrix element to error $\epsilon$ is $O(N^6m^3/\epsilon^2)$. The total cost of computing $\bf  \mathcal{C}$ is then $O(N^8m^5/\epsilon^2)$.

$\quad$\\
\noindent{\bf Computing $\bf  \mathcal{Y}$}. From Equation \eqref{eq:C} we see that to compute a vector element $y_{k\alpha}$, we need to compute terms of the forms $\langle\frac{\partial \psi_k}{\partial z_{k\alpha}}|H|\psi_\ell\rangle$, $\langle\frac{\partial \psi_{k\alpha}}{\partial z_{k\alpha}}|L_j|\psi_\ell\rangle$, and $\langle\frac{\partial\psi_k}{\partial z_{k\alpha}}|L_j^\dagger L_j|\psi_\ell\rangle$. Computing these cost $O(N^2m^2(r^2+s^4)/\epsilon^2)$. Because $ \mathcal{Y}_{k\alpha}$ contains $O(N^3)$ terms, the total cost for evaluating $ \mathcal{Y}_{k\alpha}$ is $O(N^8m^2(r^2+s^4)/\epsilon^2)$. To compute the entire vector $\bf  \mathcal{Y}$ then costs $O(N^8m^2(r+s^2)/\epsilon^2)$.





\section{Numerical example}
This section of the paper gives illustrative numerical examples of the performance of the HQC algorithm to dynamics of  open systems. For simulating of open system dynamics via outer-product method: $First$, we generate an initial state $|\psi({ z_{k\alpha}}(0))\rangle$  with a sequence of gates parameterized by unitary quantum circuit and contains set of control parameters  $\{{ z_{k\alpha}}(0)\}$to be optimized on the quantum processor. Then, we solve the differential Eq.\eqref{eq:eom} numerically via the classical computer, in which the matrices $\mathcal{C}, \mathcal{Y}, \mathcal{S}$, {\bf L}  and $ {\tau} $ in the equation are evaluated using the sequences of unitary quantum operations in quantum computer. $Second$, by using the solution, we can project parameters forward by a small-time $\delta t$  and with the iteration the parameters of $z_{k\alpha}(t)$ are determined iteratively given their initial values.
We show in the Fig. [4] a  possible implementation of quantum circuit to obtain the dynamics of an open system that includes different layers for measuring different matrices via outer-product method.

\begin{figure}[h!]
\begin{center}
\includegraphics[width=14cm]{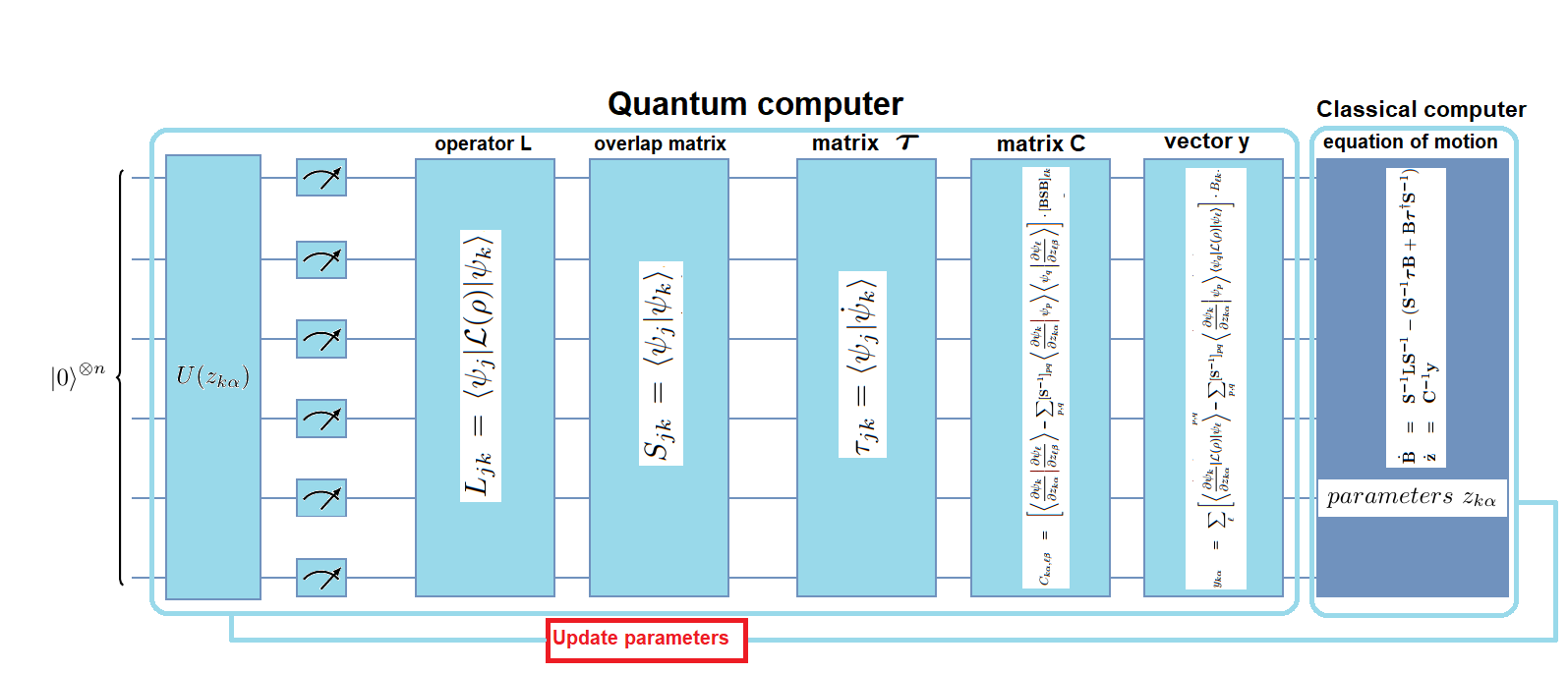}
\caption{HQC algorithm for solving quantum dynamics via the outer product method. The task of the classical computer
is to determine $z_{k\alpha}$ parameters and quantum computer evaluate quantum state.}
\end{center}
\end{figure}

To demonstrate our method, we implemented numerical examples for different open systems.\\
$Dephasing$ $noise$. For the dephasing process (with the rate $ \gamma$) that destroys the phase coherence of any superposition, we consider simulating the dynamics of the density matrix which is modeled by a Markovian master equation of

\begin{equation} \label{E20}
\dot{\rho}(t)=-i[H_s,\rho]+ \gamma \sigma_z \rho \sigma^\dagger_z - \frac{\gamma}{2}\sigma^\dagger_z \sigma_z \rho -  \frac{\gamma}{2} \rho \sigma^\dagger_z \sigma_z ,
\end{equation}
where $\sigma_z$ is pauli matrix and $H_s$ is Hamiltonian of system. By considering the $H_s=\frac{\omega_0}{2}\sigma_z$  and initial state $\rho_0=|\psi \rangle \langle \psi |$ with $|\psi \rangle =\frac{1}{\sqrt{2}}(|0\rangle +|1 \rangle)$, then evaluating the Eq.\eqref{E20} leads to time evolution of $\rho(t)$. We suppose $|\psi_1 \rangle=e^{i\sigma_z z_1}|0 \rangle$ and $|\psi_2 \rangle=e^{i\sigma_z z_2}|1 \rangle,$
and by using Eq.\eqref{eq:eom} and Eq.\eqref{eq:C} we can obtain elements of $L_{kl}$, $\mathcal{S}$ , $\mathcal{C}_{k\alpha,l\beta}$ and $\mathcal{Y}_{k\alpha}$. We use  expectation value of pauli matrix $\sigma_x$ that defined as $\langle \sigma_x \rangle =Tr(\rho \sigma_x)$ for calculation the effect of dephasing process on  phase coherence system. Output of hybrid quantum-classical algorithm compared it to exact solution  $\langle \sigma_x \rangle$ as shown in Fig. [2].

\begin{figure}[h!]
\centering
\subfloat[]{\includegraphics[width=10cm]{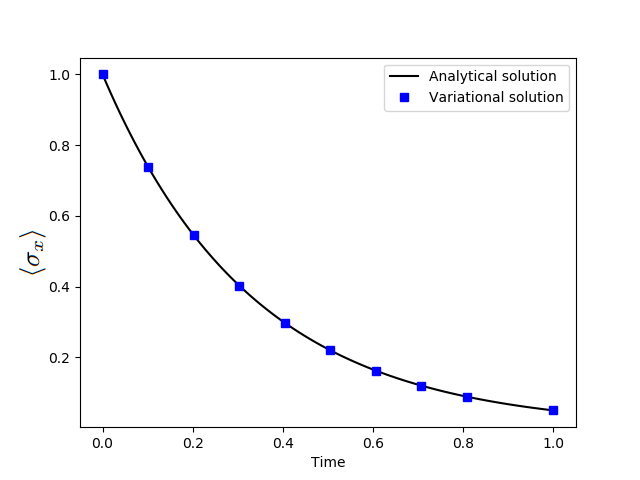}}
\caption{  Value  $\langle \sigma_x \rangle$ in the two cases exact  and hybrid method. Initial parameters are $\gamma =1.5$, $z_1=z_2=1$ and $B_{11}=B_{22}=1,\ B_{12}=B_{21}=1$ }
\end{figure}

$Dissipative$ $noise$. We consider  dissipative dynamics of the open quantum system which is described by following master equation
\begin{equation}\label{E22}
\dot{\rho(t)}=-i[H_s,\rho]+ \Gamma \sigma^- \rho \sigma^+ - \frac{\Gamma}{2}\sigma^+ \sigma^- \rho -  \frac{\Gamma}{2} \rho \sigma^+ \sigma^- ,
\end{equation}
where $\Gamma$ is dissipation rate and $\sigma^-$ and $\sigma^+$ are raising and lowering operators, respectively. Here, we suppose $H_s=\frac{\omega_0}{2}\sigma_z$ , $\rho_0=|0 \rangle \langle 0 |$ , $|\psi_1 \rangle =e^{i\sigma_z z_1}|0 \rangle $ and $|\psi_2 \rangle =e^{i\sigma_x z_2}|1 \rangle $. We investigate the effect of the dissipation process on the energy system with defining $E=Tr(\rho H)$. The result of the hybrid quantum-classical algorithm compared it to the exact solution $E$ as indicated in Fig. [3].
\begin{figure}[h!]
\centering
\includegraphics[width=10cm]{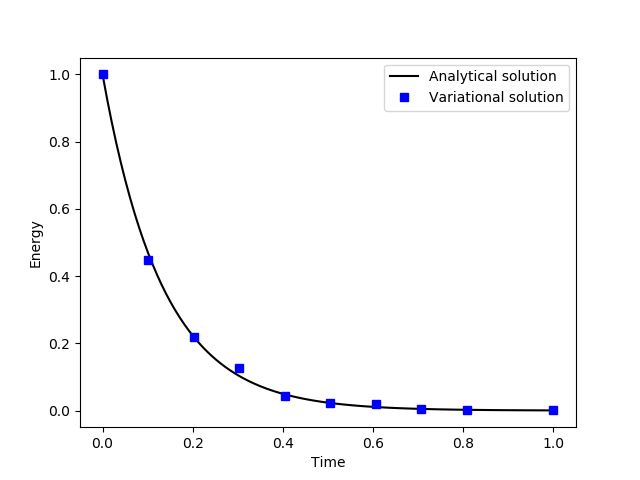}
\caption{System Energy in the two case exact  and hybrid methods. Value of initial parameters are $z_1=0$, $z_2=\pi/4$ and $ B_{11}=B_{12}=B_{21}=B_{22}=1$ and $\Gamma=7.5$}
\end{figure}
\newpage
\emph{Two- level system coupled to a photon mode} .
We consider a simple harmonic oscillator connected to the Two-level system (TLS) as a closed system that coupling to a reservoir\cite{bh}. A Hamiltonian that represents the TLS–oscillator system is given by
\begin{equation}
H=\hbar \omega_r a^\dag a +\frac{\hbar \omega_r}{2}\sigma_z + G(a \sigma^+ +a^\dag \sigma^-),
\end{equation}
where $\hbar$ is Planck’s constant, $a$ and $a^\dag$ are annihilation and creation
operators respectively, which together obey the bosonic commutation and  $G $ is  coupling strength. Here $\sigma^-$ and $\sigma^+$ are raising and lowering operators, respectively.
This is the
Jaynes–Cummings Hamiltonian \cite{jaynes} and for which the eigenenergies and
eigenstates can be obtained analytically\cite{gerry}. The master equation describes that TLS–oscillator system interaction with reservoir given with
\begin{equation}\label{E12}
\dot{\rho(t)}=-i[H,\rho]+ \frac{\gamma}{2}( 2\sigma^- \rho \sigma^+ - \sigma^+ \sigma^- \rho -  \rho \sigma^+ \sigma^-) .
\end{equation}
We can obtain oscillator
photon number the excited and ground-state populations of
the TLS by numerically solving Eq.\eqref{E12}.
In the case of quantum computing, qubits are  computational basis, so we first mapped Hamiltonian into qubits. We use the basis of harmonic oscillator eigenstates,
as these can be easily mapped to qubits \cite{som}. By considering the truncated eigenstates with the lowest $d$ energies $|s\rangle$  and
employ binary representation we can encode $|s\rangle$ to qubits\cite{mc}. Therefore
$|s\rangle=|b_{k-1}\rangle|b_{k-2}\rangle....|b_0\rangle$ where $s=b_{k-1}2^{k-1}+b_{k-2}2^{k-2}+ ... b_0 2^0$. The representation of the creation operator is
\begin{equation}\label{E13}
a^\dag=\sum_{s=0}^{d-2}\sqrt{s+1}|s+1\rangle\langle s|,
\end{equation}
and the annihilation operator can be obtained by taking the
Hermitian conjugate of $a^\dag$.  We can  map  these binary projectors to Pauli
operators
$$|0\rangle \langle 0| =\frac{1}{2}(I+\sigma_z) $$
$$ |1\rangle \langle 1| =\frac{1}{2}(I- \sigma_z)$$
$$|0\rangle \langle 1| =\frac{1}{2}(\sigma_x+ i\sigma_y)$$
\begin{equation}
  |1\rangle \langle 0| =\frac{1}{2}(\sigma_x - i\sigma_y).
\end{equation}

Now back to the first problem, here we consider.
$$|\psi_1 \rangle =e^{i\sigma_z z_1}e^{ia^\dag a z_1}|n \rangle _{h} |0 \rangle _s ,$$
and
$$|\psi_2 \rangle =e^{i\sigma_z z_2}e^{ia^\dag a z_2}|n \rangle_h |1 \rangle_s,$$
where $h$ and $s$ refer to  harmonic oscillator and TLS, respectively. By considering n=2 that is photon number  in the oscillator and using Eq.\eqref{E13} and binary representation  $|\psi_1 \rangle $ and $|\psi_2 \rangle $ can be mapped to qubit form

$$|\psi_1 \rangle =e^{i\sigma_z z_1}e^{i I\otimes I z_1 }e^{i I\otimes \sigma_z z_1 }e^{i \sigma_z\otimes I z_1 }|00\rangle |0\rangle,  $$
\begin{equation}
|\psi_2 \rangle =e^{i\sigma_z z_2}e^{i I\otimes I z_2 }e^{i I\otimes \sigma_z z_2 }e^{i \sigma_z\otimes I z_2 }|00\rangle |1\rangle.
\end{equation}
Here we obtain oscillator photon number that define $\langle N \rangle = Tr(\rho N)$  where $ N=a^\dag a $. The numerical results shows in the Fig. [4].
\begin{figure}[h!]
\centering
\includegraphics[width=10cm]{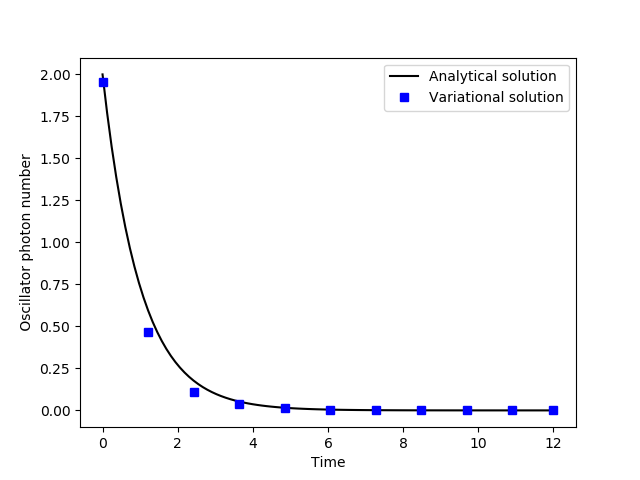}
\caption{Oscillator photon number in two cases exact  and hybrid method. Value of initial parameters are $z_1=0$, $z_2=0 $ and $ B_{11}=B_{12}=B_{21}=B_{22}=1$ and $\gamma=10$, $G=2$. Here the somewhat deviation is at only two
points, and the other points have compatible with the exact calculation and value error is $(2\pm 2)\times 10^{-2}$.The error of variational data
 come from the system calculation errors depend on number
of repetitions and initial value of parameters.   }
\end{figure}

\emph{Linear vibronic coupling model}.
We  consider the  linear vibronic coupling model of crossing
surfaces. The system contains two electronic states donor $|D\rangle$ and  acceptor $|A\rangle$, that coupled through
two nuclear \cite{Joubert-Doriol2015Problem-freeSystems,dom,d}. A Hamiltonian that describes
this system given by
\begin{equation}
H=\sum_{i=1} ^2 \frac{\omega_i}{2}(p^2_i+x_i^2)[|D\rangle \langle D| + |A\rangle \langle A|] - d x_1[|D\rangle \langle D| -|A\rangle \langle A|]+c x_2[|D\rangle \langle A| + |A\rangle \langle D|],
\end{equation}
where  $\omega_i$ is the frequency of the coordinate $x_i$, and d and c are coupling constants. The dissipative part is described through the bilinear coupling of the system coordinates $x_i$ with coordinates of harmonic bath oscillators. The master equation that describes the effect of dissipative dynamics given in  Lindblad form
\begin{equation}
\mathcal{L}(\rho)= -i[H,\rho]+h_1\sum_i^2 (2L_i \rho L_i^\dag -L_i^\dag L_i \rho -\rho L^\dag_i L_i) +h_2\sum_i^2 (2L_i \rho L_i^\dag -L_i^\dag L_i \rho -\rho L^\dag_i L_i),
\end{equation}
 where $h_1$ and $h_2$ are  coupling constants, and $L_i$ are Lindblad operators that defined as
\begin{equation}
L_1=(a_1-\frac{d}{\omega_1 \sqrt{2}} )|D\rangle \langle D| + (a_1+ \frac{d}{\omega_1 \sqrt{2}}) |A\rangle \langle A|  ,\quad L_2=a_2(|D\rangle \langle D| + |A\rangle \langle A|),
\end{equation}
where $a_j=(x_i +i p_j)/\sqrt{2}$ are annihilation operators. In the hybrid method, we need  to transform  Hamiltonian to qubit form. To this end by setting
the system contains two electronic states as a two-level system and
$|D\rangle = |0\rangle$ and $|A\rangle = |1\rangle$ and $a^\dag a =p^2+x^2$ we obtain
\begin{equation}
H=\sum_{i=1} ^2 \frac{\omega_i}{2}(a^\dag_i a_i) - d (a^\dag_1 +a_1)\sigma_z+c (a^\dag +a) \sigma_x,
\end{equation}
This Hamiltonian describes the two harmonic oscillators is coupled with a  two-level system (TLS) and  the dissipative part is introduced coupling of
the system with reservoir. The schematic of this system as shown in Fig. [5].
\begin{figure}[h!]
\centering
\includegraphics[width=10cm]{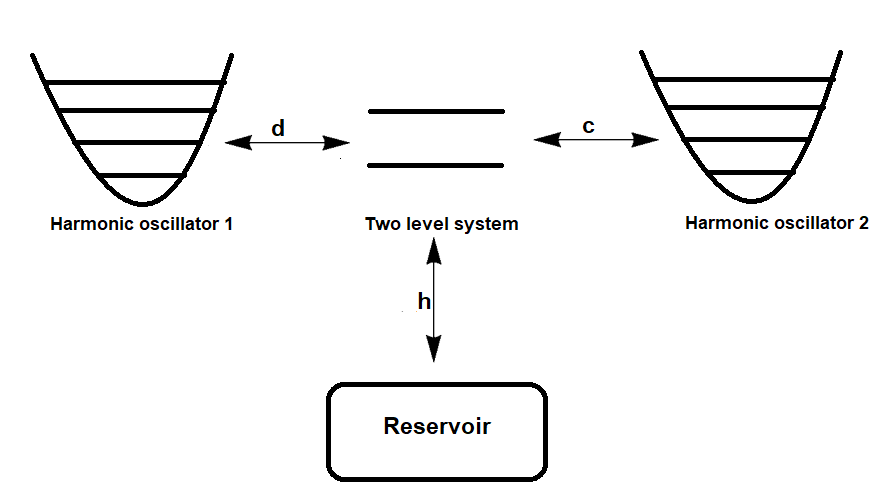}
\caption{A schematic of the  system described in Example 4. The two harmonic oscillators are coupled a TLS with strenght d and c  and coupled to reservoir whit  strength h }
\end{figure}
Similar Example 3 we employ  binary representation for mapping Hamiltonian to qubit form. Note that, also Lindblad operators mapping to qubit form.
So by considering
$$|\psi_1 \rangle =e^{ i\sigma_z z_1}e^{i \sum_{j=1}^2 I^j\otimes I^j  z_1 }e^{i \sum_{j=1}^2 I^j\otimes \sigma_z^j z_1 }e^{i \sum_{j=1}^2 \sigma_z^j\otimes I^j z_1 }|00\rangle |00\rangle |0\rangle,  $$
\begin{equation}
|\psi_2 \rangle =e^{ i\sigma_x z_2}e^{i \sum_{j=1}^2 I^j\otimes I^j  z_2 }e^{i \sum_{j=1}^2 I^j\otimes \sigma_z^j z_2 }e^{i \sum_{j=1}^2 \sigma_z^j\otimes I^j z_2 }|00\rangle |00\rangle |1\rangle.
\end{equation}

Here we obtain  expectation value of Pauli matrix $\sigma_z$ that given by $\langle \sigma_z \rangle =Tr(\rho \sigma_z)$ for study losing energy of TLS. The exact and hybrid solution comparing and results show in the Fig. [5].
\begin{figure}[h!]
\centering
\includegraphics[width=10cm]{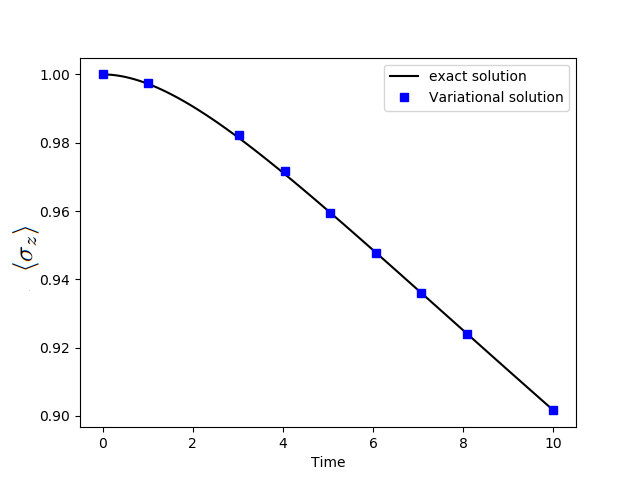}
\caption{Value $\sigma_z$ in the two cases exact and hybrid method. Initial parameters are $h_1=1,\ h_2=0$, $\omega_1=\omega_2=0.007$ ,$d=0.05,\ c=0.04$ and $z_1=z_2=0$ and $ B_{11}=B_{22}=B_{12}=B_{21}=1$  }
\end{figure}

\newpage

\section{Conclusion}

Using a novel quantum gate model, we have developed a HQC algorithm in near-term quantum resources to dynamics of open systems described by a master equation with generator $\mathcal{L}$ in Lindblad form. We have used the time-dependent variational principle to density matrix formalism involves minimization of the Frobenius norm of the error. We compare the analytical and numerical results obtained from the exact solution of the master equation with the HQC algorithm given by outer-product methods by several different examples for fermionic and bosonic environments which are in good agreement with each other.
In the future, we will extend these results to the non-Markovian dynamics of open systems.

{\bf Acknowledgements:}
The authors would like to thank Yudong Cao, Peter D. Johnson and Jhonathan Romero Fontalvo for useful discussions
during the preparation of this work.



\bibliographystyle{plainnat}
\bibliography{vosq}

\end{document}